\documentstyle[12pt,aps,epsf]{revtex}
\begin{document}
\preprint{DOE/ER/40427-05-N64}
\draft
\title{
{\vskip 30pt}
THE ``HOLE" IN HELIUM}
\author{L. Wilets, M. A. Alberg\footnote{permanent address: Department of 
Physics, Seattle University, Seattle WA 98122, USA}}
\address{Department of Physics, Box 351560, University of Washington, 
Seattle, WA 98195-1560, USA}
\author{J. Carlson}
\address{Theoretical Division, Los Alamos National Laboratory, Los Alamos, 
New Mexico 87545, USA}
\author{W. Koepf}
\address{School of Physics and Astronomy\\
         Raymond and Beverly Sackler Faculty of Exact Sciences\\
         Tel Aviv University, 69978 Tel Aviv, Israel}
\author{S. Pepin and Fl. Stancu}
\address{Universit\'{e} de Li\`ege\\
         Institut de Physique B.5\\
         Sart-Tilman, B-4000 Li\`ege 1, Belgium}
\maketitle
\begin{abstract}

The measurement and analysis of electron scattering from $^3$He
and $^4$He by Sick and collaborators reported 20 years ago remains a matter
of current interest.  By unfolding the measured free-proton charge 
distribution, they deduced a depression in the central point nucleon
density, which is not found in few-body calculations based on realistic
potentials.  We find that using wave functions from such calculations we can
obtain good fits to the He charge distributions under the assumption that 
the
proton charge size expands toward the center of the nucleus.  The
relationship to 6-quark Chromo-Dielectric Model calculations, is discussed.  The
expansion is larger than than the predictions of
mean field bag calculations by others or our CDM
calculations in the independent pair approximation.  There is interest here
in the search for a ``smoking gun" signal of quark substructure.

\end{abstract}
\section{INTRODUCTION}

The charge distribution of nuclei has been the subject of experimental
studies for more than forty years.  Electron scattering and muonic atoms
now provide detailed descriptions of the full range of stable, and many
unstable nuclides. Unique among the nuclides are the isotopes $^3$He and
$^4$He because they exhibit a central density about twice that of any
other nuclide. There is a long-standing apparent discrepancy between the
experimentally extracted charge distributions and detailed
theoretical structure calculations which include only nucleon degrees
of freedom. 

McCarthy, Sick and Whitney\cite{mc,sick} performed electron scattering 
experiments on these isotopes up to momentum transfers of 4.5 fm$^{-1}$
yielding a spatial resolution of $~0.3$ fm.  They extract a ``model 
independent" charge distribution, which means that their analysis of the data 
is not based upon any assumed functional form for the charge distributions.
Their charge distributions are shown in Fig. 1.  Taken alone, they  do not
appear 
to be extraordinary.  However, using a finite proton form factor, which 
fits the experimentally measured 
rms radius of about 0.83 fm, they unfolded the 
proton structure from the charge distributions to obtain the proton point 
distributions.  For both isotopes there is a significant 
central depression of about 30\% extending to about 0.8 fm.  
Sick\cite{sick} also presented results where relativistic and meson effects 
are included.  
These are shown for $^4$He in Fig. 2.
One note of caution here is that it is not possible to subtract these
effects from the experimental data in a completely model-independent way.

\begin{figure}[b]
        \centering
        \leavevmode
        \epsfxsize=4in
        \epsfysize=5.5in
        \epsffile{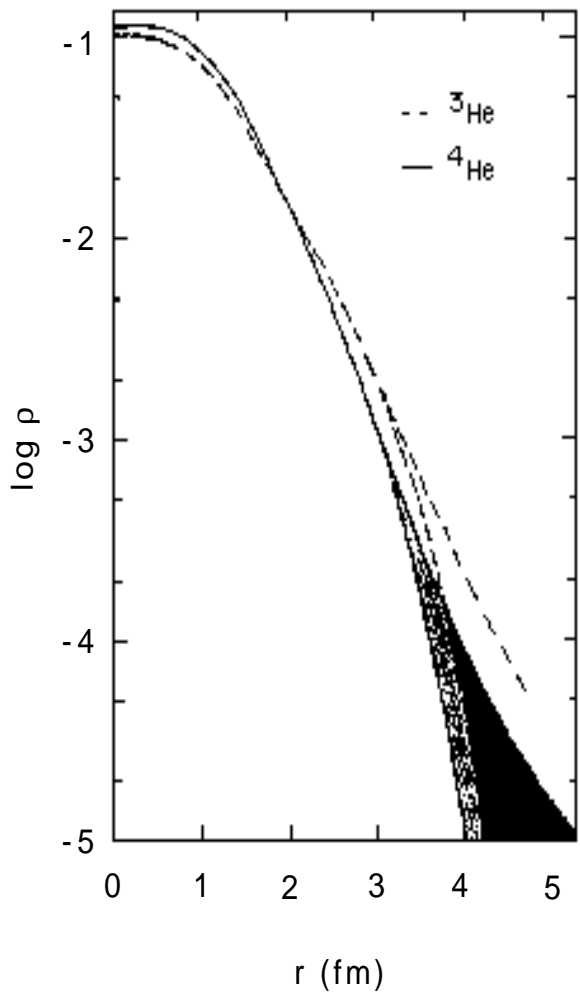}
\noindent\caption{Model-independent charge distributions extracted 
from experiment.  Reproduced from McCarthy {\it et al.}[1].
\label{fig1}}
\end{figure}

\begin{figure}
        \centering
        \leavevmode
        \epsfxsize=5in
        \epsfysize=5in
        \epsffile{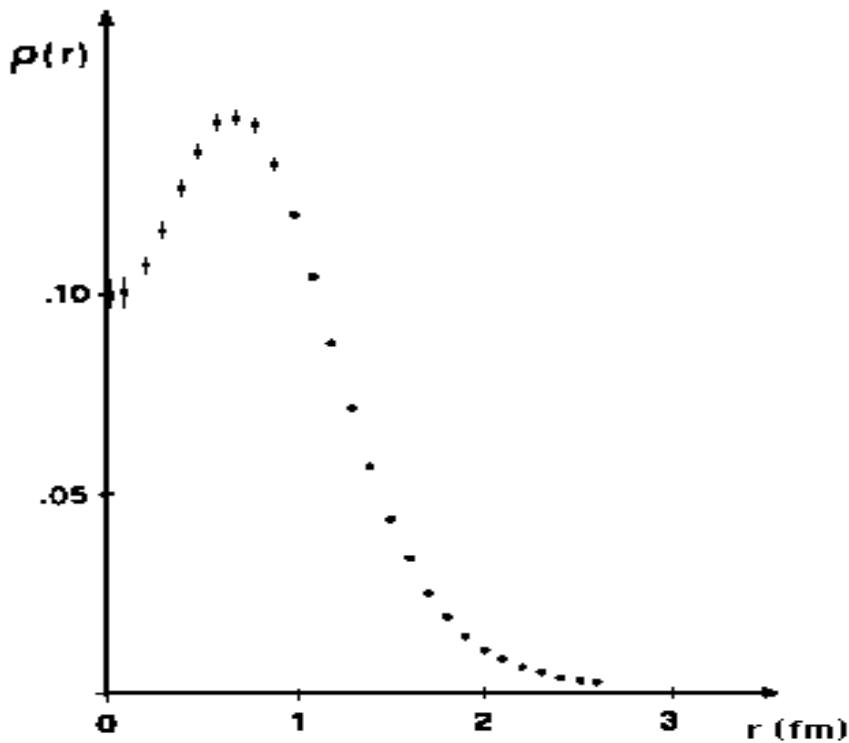}

\noindent \caption{Point-proton density distributions obtained by 
unfolding the finite free proton form factor, allowing for meson exchange 
corrections and relativistic effects.  Reproduced from Sick[2].
\label{fig2}}
\end{figure}

One might assume that such a central depression is to be expected because 
of the short-range repulsion of the nucleon-nucleon interaction.  So far, 
this is 
not borne out by numerous detailed theoretical calculations, none of which 
finds a {\it significant} central depression, certainly not of the above 
magnitude. Relatively smaller central depressions are found in Green's 
function Monte Carlo (GFMC)
calculations of the alpha particle for realistic models of the two- and
three-nucleon interaction.  (see Fig. 4 below.)

The status of theoretical structure calculations through mass number 4 is 
very satisfactory at present.  Given any assumed interaction, the few body 
problem can be solved to within tenths of an MeV in energy and the wave
function can be calculated to a precision better than that required for the 
present discussion.

In using a nuclear wave function to construct a charge distribution, one
must assume a nucleon charge density and the possibility of
meson exchange contributions.  
While the meson exchange contributions
in the transverse channel are well-constrained (at least at moderate
momentum transfer) by current conservation, no such constraint is 
available in the longitudinal channel. Indeed, meson exchange current
contributions are of relativistic order and hence one must be careful
when interpreting them with non-relativistic wave functions.

Given these caveats, it is possible to reproduce reasonably well the 
longitudinal form factors of 3- and 4-body nuclei within a 
nucleons plus meson-exchange model.\cite{currents,ff} 
The current and charge
operators are constructed from the N-N interaction and required to
satisfy current conservation at non-relativistic order. The
resulting meson-nucleon form factors are quite hard, essentially 
point-like.\cite{currents} 
This raises the possibility of explaining the
form factors in quark or soliton based models, which would describe the
short-range two-body structure of the nucleons in a more direct
way than is available through meson exchange current models.
For example, see the discussion of a model by Kisslinger {\it et al.}\cite{kis} 
below.

We present here a possible explanation of the electric form factors which 
does not involve a hole in the point-proton density distribution, 
but rather is consistent 
with theoretical few body calculations.  It involves the variation of the 
proton charge size as a function of density, or as a function of
nucleon-nucleon separation.  This is not depicted as an average
`swelling' of the nucleon, but as a result of short-range dynamics in the 
nucleon-nucleon system.  The results presented here are preliminary but 
encouraging.

\section{QUARK SUBSTRUCTURE OF NUCLEI AND NUCLEONS}

Within the context of soliton models, there have been numerous 
calculations of the nucleon size in nuclear media.  Most of these involve 
immersion of solitons in a uniform (mean) field generated by other 
nucleons.  Another approach has been the 3-quark/6-quark/9-quark bag
models, which has been applied to various nuclear properties, including the 
EMC effect.  It has been applied by Kisslinger et al.\cite{kis} to the He 
electric form factors with some success.

In a series of papers, Koepf, Pepin, Stancu and 
Wilets\cite{statics,emc,dynamics} have studied the 
6-quark substructure of the two-nucleon problem, and in particular obtain 
the variation of the quark wave functions with inter-nucleon separation.
Contrary to previous expectations, the united 6-quark cluster does not 
exhibit a significant decrease in the quark momentum distribution function
in spite of an 
increase in the volume available to the individual quarks.\cite{emc}  
This is due to 
configuration mixing of higher quark states.   Such a momentum
decrease was proffered as an explanation of the EMC effect.  However, the 
united cluster does have approximately twice the volume of confinement 
of each 3-quark 
cluster, and the quarks extend to a volume nearly three times that of the 
3-quark clusters, again enhanced by configuration mixing of excited states.

In Fig. 3 we exhibit the proton rms charge radius $r_p$ 
extracted as follows from
the calculations of 
Pepin {\it et al.}\cite{dynamics}:  the abscissa 
gives the effective nucleon-nucleon separation $r_{NN}$ 
obtained by the Fujiwara 
transformation; the soliton-quark structure is a 6-quark deformed 
composite.  The proton rms radius is defined to be
\begin{equation}
r_p=\sqrt{<r^2>-r_{NN}^2/4}
\end{equation}
where the quark density used in calculating $<r^2>=\int \rho_q r^2 d^3r$ 
is the six-quark density normalized to unity.

\vbox{
\begin{figure}
\vskip -.5truein

        \centering
        \leavevmode
        \epsfxsize=4.3in
        \epsfysize=5.1in
        \epsffile{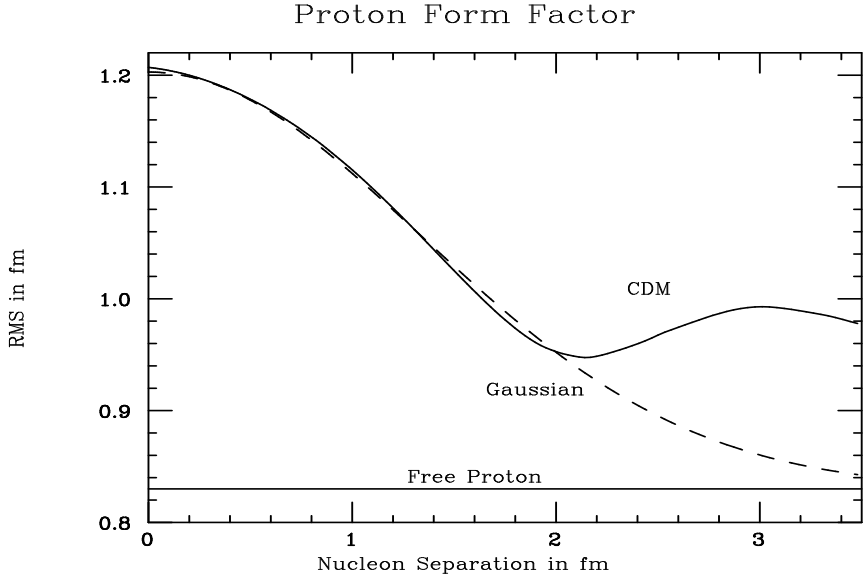}
\vskip -.5truein
        \caption{Proton rms charge radius as a function of inter-nucleon 
separation.  The dashed line is a gaussian approximation, normalized to the 
free value.}
        \label{Figure_3}
\end{figure}
}
 
For well separated solitons, the $r_{NN}$ is just the separation of the soliton 
centers and $r_p=0.83$ fm as indicated by the horizontal line.  Large 
deformations (near separation) are difficult to calculate so that the 
figure does not reproduce well the separation region.  Shown also in the 
figure is a gaussian approximation fitted to $r_{NN}=0$, $r_{NN}=1$ fm, and 
the asymptotic region, see Eq. (5) below, with $A=0.45,\ s=0.92$ fm.
Then the charge distribution due to two-body correlations is
\begin{eqnarray}
&f_p(\bbox{r_i,r_j;r})=\Big\{\delta_{ip}
\ \exp\big[-|\bbox{r-r_i}|^2/b^2(r_{ij})\big]\nonumber\\
&+\delta_{jp}\ \exp\big[-|\bbox{r-r_j}|^2/b^2(r_{ij}\big])\Big\}/2\,
\pi^{3/2}b^3(r_{j})
\end{eqnarray}
where we assumed and indicate explicitly that $b$ is a function of the
distance $r_{ij}$ between the nucleons i and j, as we expand upon later, and
that the proton and neutron functions are the same.  Here ``$p$" stands for 
``proton" and the Kronecker deltas pick out protons among $i$ and $j$.

Using the independent pair approximation (IPA) and Eq. (2)
we find the charge distribution by employing a two-body correlation function
$\rho_2(\bbox{r_i,r_j})$,
\begin{equation}
\rho_{ch}(r)=\sum_{i<j}\int d^3r_i\int d^3r_j\,\rho_2(\bbox{r_i,r_j})
f_p(\bbox{r_i,r_j;r})/3\,.
\end{equation}
There are six pairs $(i,j)$.  Each proton appears three times; Hence the 
factor 1/3.
\section{PHENOMENOLOGICAL DENSITY-DEPENDENT PROTON FORM FACTOR}

To obtain some qualitative feeling for the expansion of the proton charge 
size with nucleon density, we assume a proton form factor $f_p$ (differing 
from Eq. (2)) with a size that
depends simply on the local density and hence on the distance from the 
center of the nucleus $(r')$.  Then
\begin{equation}
\rho_{ch}(r)=\int d^3r' \rho(r') f_p(\bbox{r,r'})\,,
\end{equation}
where $\rho$ is the (theoretical) point proton density

\begin{equation}
f_p(\bbox{r,r}') = \exp \big[-|\bbox{r-r}'|^2/b^2(r')
\big]\ /\pi^{3/2}b^3(r')
\end{equation}
and we choose
\begin{equation}
b(r')=b_0 [1+Ae^{-r'^2/s^2}]\,,
\end{equation}
with $b_0=\sqrt{2/3}$ 0.83 fm, the free proton value.
A fairly good fit to the data was found with $A=0.45$ and $s=0.65$ fm 
corresponding to $b(0)$ equal to the central value given in Fig. 3.  The 
best fit, with only slightly better $\chi$-squared, 
was obtained with $A=2.10$ and $s=0.13$ fm, which does 
not seem to be reasonable, in that the $A$ is too large and the $s$ too small.

\vbox{
\begin{figure}[b]
\vskip -1.0truein

        \centering
        \leavevmode
        \epsfxsize=4.3in
        \epsfysize=5.1in
        \epsffile{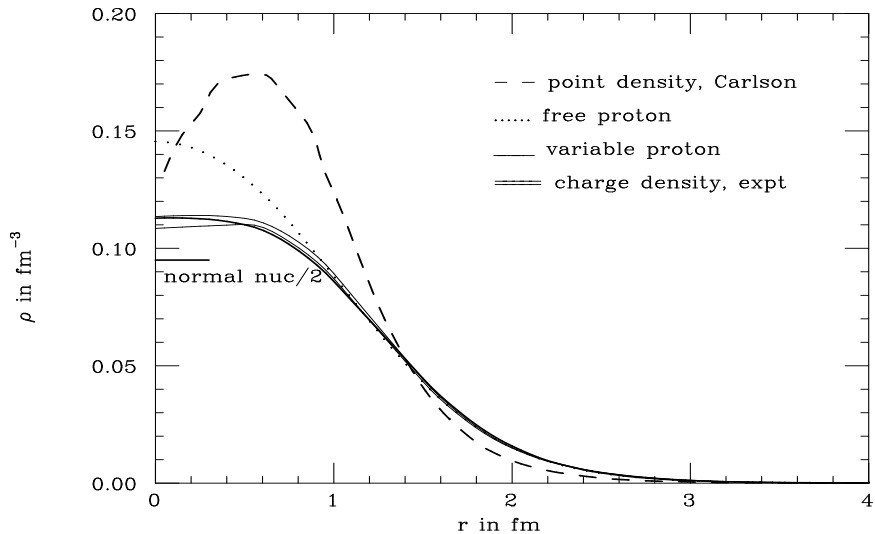}
\vskip -1truein
        \caption{$^4$He density distributions including a fit with a 
variable proton charge size.  The dashed curve labelled ``Carlson" is based 
on a Green's function Monte Carlo calculation [4]}
        \label{Figure_4}
\end{figure}
}
\section{THE INDEPENDENT PAIR APPROXIMATION}

      In the spirit of the independent pair approximation, the charge 
distribution was calculated using Eq. (3) with $\rho_2$ the two-particle 
correlation function\cite{ff}.  The proton size parameter 
$b(r_{12})$ was first taken from the gaussian fit to the 
calculations of Pepin {\it et al.}\cite{dynamics}.  The improvement over the 
free constant proton size, as shown in Fig. 5 (dot-dash) was small.  

A phenomenological fit to the 
data was made with a parameterized $b$ given by
\begin{equation}
b(r_{12})=b_0 [1+Ae^{-(r_{12}/s)^n}]\,,
\end{equation}
where $n=2$ is of the gaussian form.  $n>2$ yields a sharper transition.
Indeed, $n\to\infty$ yields a step function.  Recall that the model of 
Kisslinger {\it et al.} corresponds to a step function.

We examined $n$ = 2, 4 and 6.  Although the $A$ and $s$ were different in 
each case, the quality of fits were very similar.  The corresponding best 
fit values of $(A,\ s,\ n)$ for the three $n$'s were (2.185, 0.883, 2), 
(0.976, 1.245, 4), (0.774, 1.34, 6).  In Fig. 5 we show the results for $n$=2
since the others are indistinguishable to the eye.
\vbox{
\begin{figure}
\vskip -1truein

        \centering
        \leavevmode
        \epsfxsize=4.3in
        \epsfysize=5.1in
        \epsffile{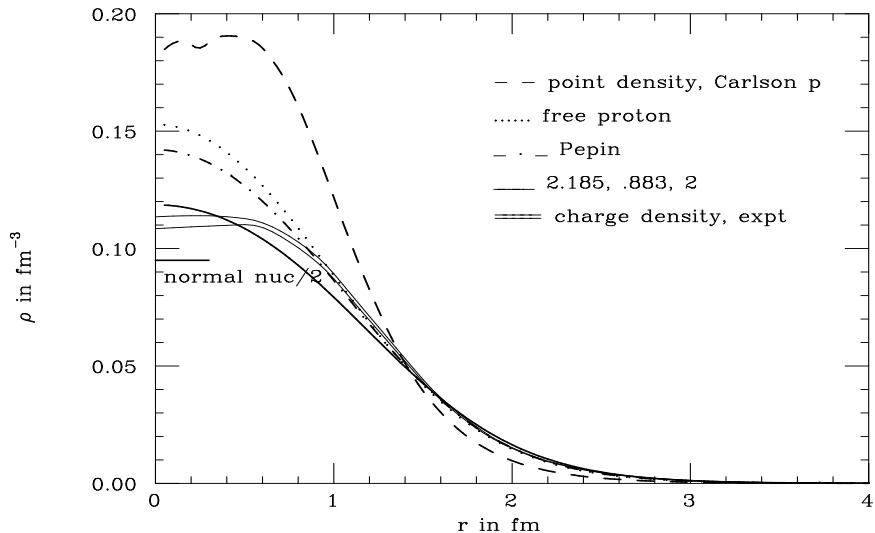}
\vskip -1truein
        \caption{$^4$He density distributions constructed from variational 
point densities and two-body correlation functions in a parameterized 
variational calculation by Carlson {\it et al.}[4].  
The curve labelled ``Pepin" uses the gaussian 
fit of Fig. 3, based on the calculations of Pepin {\it et al.}[8].
The solid curve is a phenomenological fit as described in 
the text}
        \label{Figure_5}
\end{figure}
}

\section{CONCLUSIONS}

We have obtained a phenomenological fit to the $^4$He charge distribution 
by assuming a proton size which increases with increasing density.  More 
specifically, we minimize the charge distribution $\chi$-squared using
a two-parameter gaussian function of $r$, the distance from the center-of 
mass.

We would like to identify the variable proton size with the 
structure function of Fig. 3 derived
from 6-quark $N$-$N$ studies in the spirit of 
the independent pair approximation.  Fairly good 
agreement with experiment was obtained with a phenomenological 
parameterization of the proton size function.

The inadequacy of the previous calculation might be due to

$\bullet$ A constant confinement volume was assumed for the six-quark 
structure as a function of deformation.  It may be that the intermediate 
volume (between separated and united clusters) is larger.

$\bullet$ The independent pair approximation may be invalid at the high 
densities of the central region.

$\bullet$ Meson effects should be recalculated  using the quark structure 
functions given (say) by the six-quark IPA model.

Items 1 and 3 are topics for further investigation. In addition,
one must study the predictions of such models for quasi-elastic scattering.
In the quasi-free regime, nucleon models produce a good description
of the data as long as realistic nucleon interactions, including charge
exchange, are incorporated in the final-state interactions.\cite{inclusive}
Unlike the charge form factor, two-body charge operators are expected to play a
much smaller role here [4].  The combination of the two regimes
provides a critical test for models of structure and dynamics in light nuclei.

\acknowledgments

This contribution is dedicated to Prof. Walter Greiner on the occasion of 
his sixtieth birthday.

We wish to thank C. Horowitz for valuable discussions.
This work is supported in part by the U. S. Department of 
Energy.

\end{document}